\documentstyle[11pt,newpasp,twoside,epsf]{article}
\begin{document}
\title{The REFLEX Cluster Survey: Probing the Mass Distribution in the
 Universe}
 \author{Luigi Guzzo}
\affil{Osservatorio Astronomico di Brera, Milano/Merate, Italy}
\author{Hans B\"ohringer}
\affil{Max-Planck-Institut f\"ur extraterr. Physik, Garching, Germany}
\author{Chris A. Collins}
\affil{Liverpool John Moores University, Liverpool, U.K.}
\author{Peter Schuecker}
\affil{Max-Planck-Institut f\"ur extraterr. Physik, Garching, Germany}
\author{\& the REFLEX Team$^1$
}
\affil{}

\begin{abstract}
We summarize some of the major results obtained so far from the REFLEX
survey of X-ray clusters of galaxies, concentrating on the latest
measurements of the cluster X-ray luminosity function and two-point
correlation function.  The REFLEX luminosity function provides the
most homogeneous census of the distribution function of masses in the
local Universe, representing a unique zero-redshift reference quantity
for evolutionary studies.  On the other hand, the observed clustering
of REFLEX clusters is very well described by the correlation function
of a low-$\Omega_M$ CDM model.  Also, the bidimensional correlation
map $\xi(r_p, \pi)$ shows no stretching along the line of sight,
indicating negligible spurious effects in the sample, with at the same
time a clear compression of the contours as expected in the presence
of coherent motions. 
\end{abstract}

\section{The REFLEX Survey}

The REFLEX\footnote{The REFLEX Team includes: H. B\"ohringer (MPE), L. Guzzo
(OAB), C.A. Collins (LJMU), P. Schuecker (MPE), G. Chincarini (OAB),
R. Cruddace (NRAL), A.C. Edge (Durham), S. De Grandi (OAB),
D.M. Neumann (Saclay), T. Reiprich (MPE), S. Schindler (LJMU),
P.A. Shaver (ESO), W. Voges (MPE)} (ROSAT-ESO Flux-Limited X-ray)
cluster survey is currently the largest sample of X-ray selected
clusters of galaxies from the ROSAT All-Sky Survey with 1) a
statistically homogeneous and fairly well-understood selection
function; 2) measured redshifts.  The survey contains 452 clusters
over the southern celestial hemisphere ($\delta<2.5^\circ$), at
galactic latitudes $|b_{II}|>20^\circ$ and is more than 90\% complete
to a flux limit of $3 \times 10^{-12}$ erg s$^{-1}$ cm$^{-2}$ (in the
ROSAT band, 0.1--2.4 keV). X-ray fluxes and source extensions were
re-measured from the RASS with a dedicated algorithm, thus avoiding
the limitations of the standard analysis software in the
characterisation of extended sources.  The details of the whole
identification process and a critical discussion of potential sources
of incompleteness are presented in B\"ohringer et
al. (2001a). Redshifts for all but 3 REFLEX clusters have been measured
during a long Key Programme (1992-2000) using ESO telescopes
(e.g. Guzzo et al. 1999), that collected more than 3500 galaxy
redshifts over almost 500 X-ray targets.
\begin{figure}
\plotfiddle{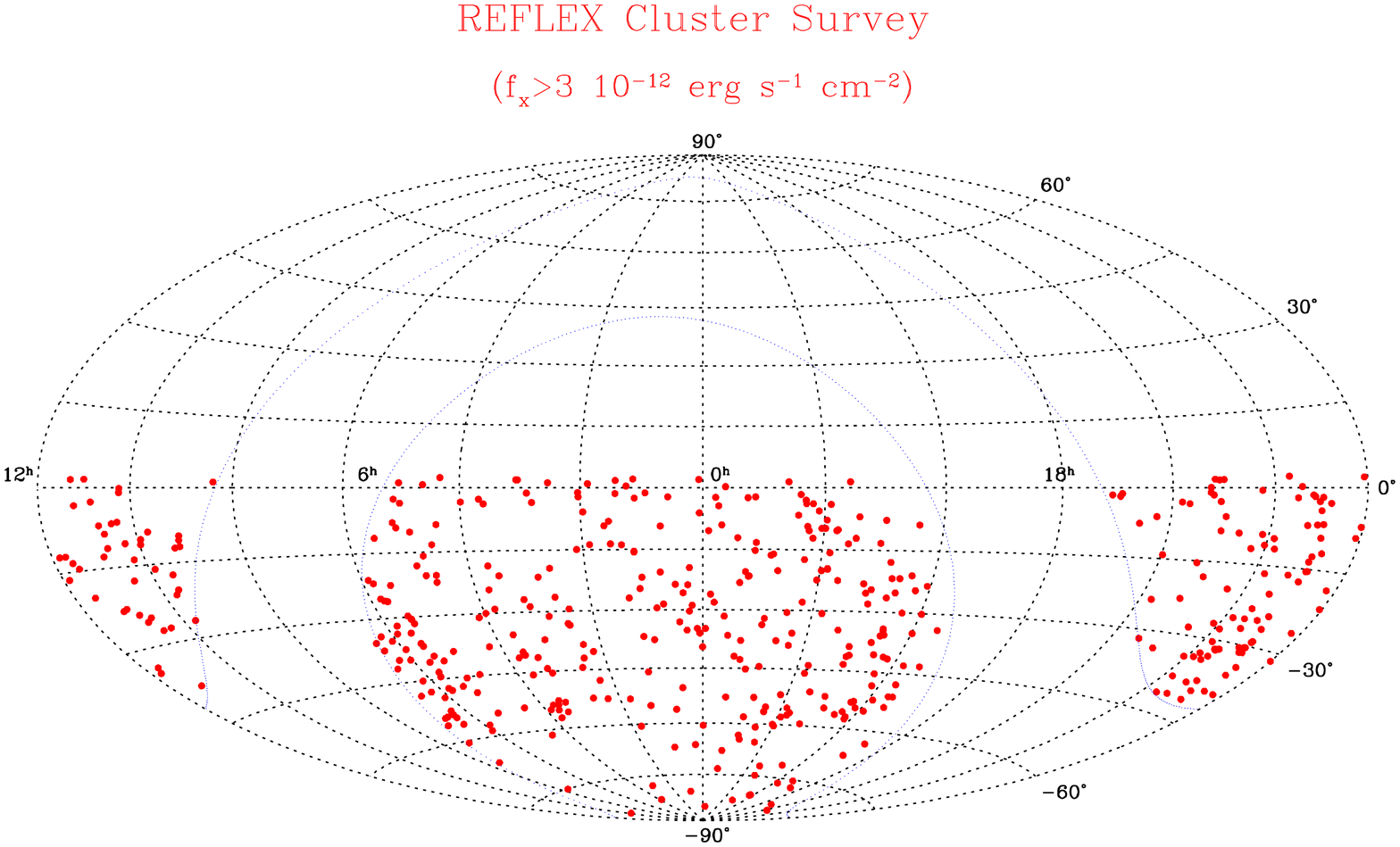}{6.0cm}{0}{40}{40}{-150}{-40}
\caption{The sky distribution of REFLEX clusters.  The dotted
band marks the region of the galactic plane ($|b_{II}|>20^\circ$)
which is excluded from the survey.}
\label{aitoff}
\end{figure}
Fig.~1 shows the
distribution on the sky of the 452 clusters in the REFLEX survey,
centered on the South Galactic Cap region (the largest contiguous area
covered), while Fig.~2 plots their X-ray luminosity against 
their redshift.  Given the survey flux limit (defined by the lower boundary
of the point distribution), at large redshifts 
we are allowed to detect only the very bright, massive
clusters.
On the other hand, the very large solid 
angle of the survey allows for an extremely large volume to be
explored (4.24 steradians, corresponding to $8.7\times 10^8$
h$^{-3}$ Mpc$^3$ out to $z=0.3$ in an $\Omega_M=0.3$,
$\Omega_\Lambda=0.7$ cosmology), such that 
these extremely rare objects on the tail of the 
cluster X-ray luminosity function have a significant probability to be
detected.  It is not by chance, in fact, that the most luminous 
cluster known is still that discovered by the REFLEX survey in 1994,
i.e. RXCJ1347.4-1144 (Schindler et al. 1995).  Another example is
shown in Fig.~4.    Such a large volume is
also ideal for sampling the very large modes of density
fluctuations in the Universe (Schuecker et al. 2001).  From the
distribution of Fig.~2 one can compute the mean density of clusters as
a function of distance, which indicates a very good completeness
(i.e. constant density) out to at least $z=0.2$, possibly above.   
It is also
evident how one can use the REFLEX survey to select sub-samples of
clusters within well-defined ranges in $L_X$ (i.e. mass) and
redshift.  A number of follow-up studies of this kind are indeed
ongoing (see B\"ohringer et al. 2001b for an overview), and more
will certainly follow when the catalogue is released in
early 2002.
\begin{figure}
\plotfiddle{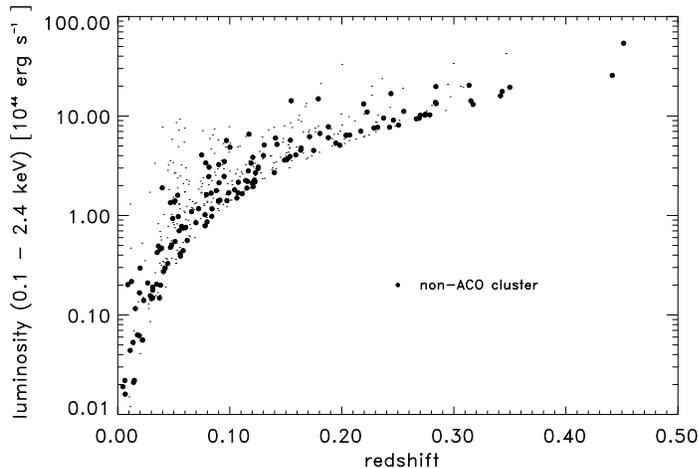}{6.0cm}{0}{50}{50}{-160}{-180}
\caption{The distribution of X-ray luminosities (ROSAT band)
as a function of redshift for all REFLEX clusters, with non-Abell
clusters marked by filled circles.  The two highest-redshift
clusters are RXCJ1206.2-0848 at $z=0.441$ (whose RGB image is shown in
Fig.~4) and RXCJ1347.4-1144, the most X-ray luminous
cluster currently known, at $z=0.452$ (Schindler et al. 1995). } 
\label{Lx_z}
\end{figure}
The filled circles in Fig.~2 represent 142 clusters which do not
appear in the visually-selected Abell/ACO catalogue (Abell 1958;
Abell et al. 1989).  It is interesting to see that these clusters are
distributed basically at any redshift.  This is a demonstration of how
the ``richness'' criterion (the reason why most of these objects did not
enter the Abell selection) is a poor indicator of the cluster mass,
and how incomplete in mass a purely visually-selected cluster
survey could be.  We have also performed a direct X-ray
analysis of the RASS data at all Abell-ACO cluster positions, measuring
their X-ray flux.  This gave the rather encouraging result that the
REFLEX survey in fact detects {\it all} Abell-ACO clusters 
within its flux and area boundaries\footnote{Even more, the REFLEX
survey misses only 1 of the so-called
Supplementary Abell clusters i.e. the extension to the main ACO catalogue,
where those objects that did not meet all original criteria while
still looking as {\it bonafide} clusters were listed.}.

\section{The Cluster X-ray Luminosity Function}

%
%
%
\begin{figure}
\plottwo{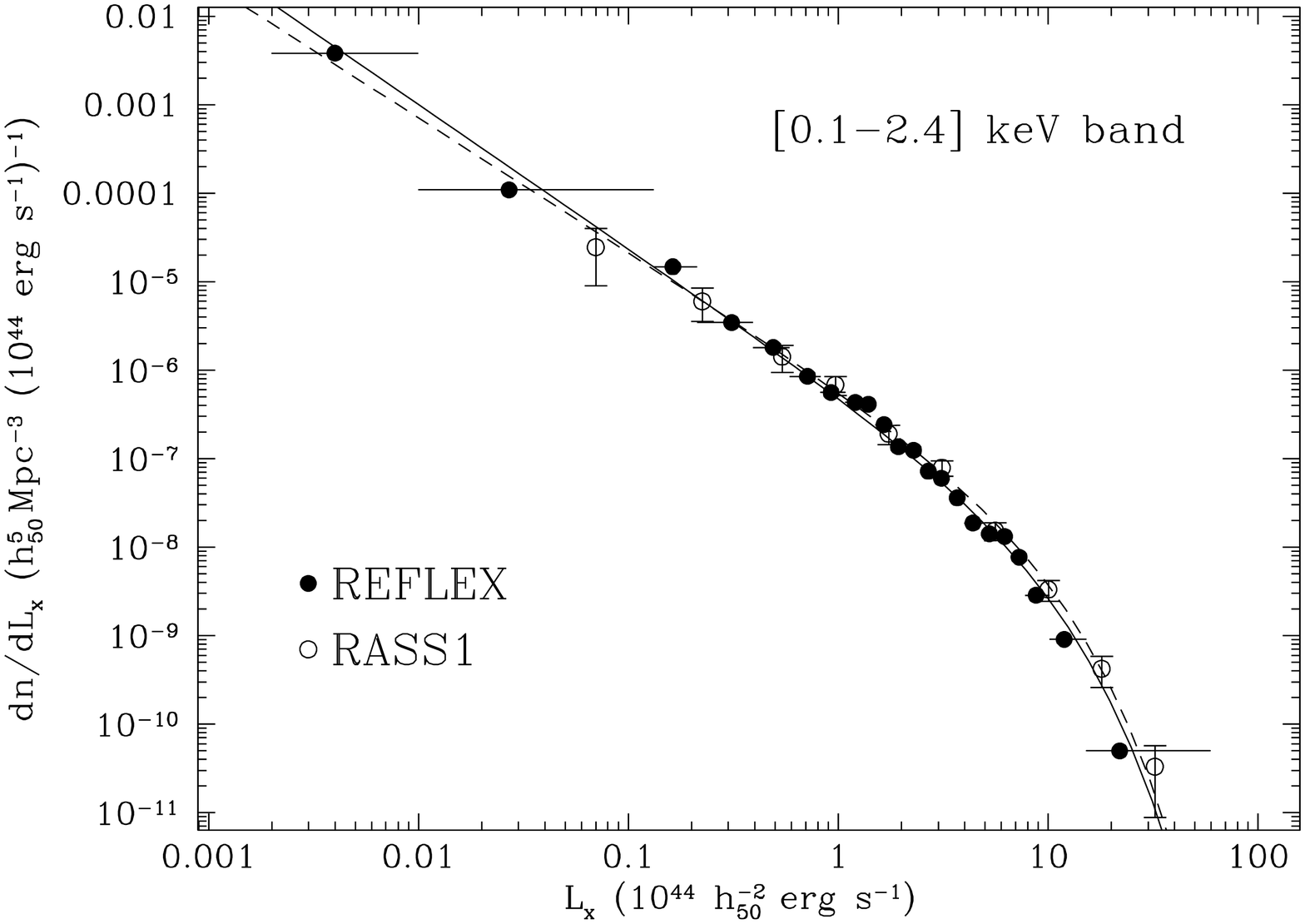}{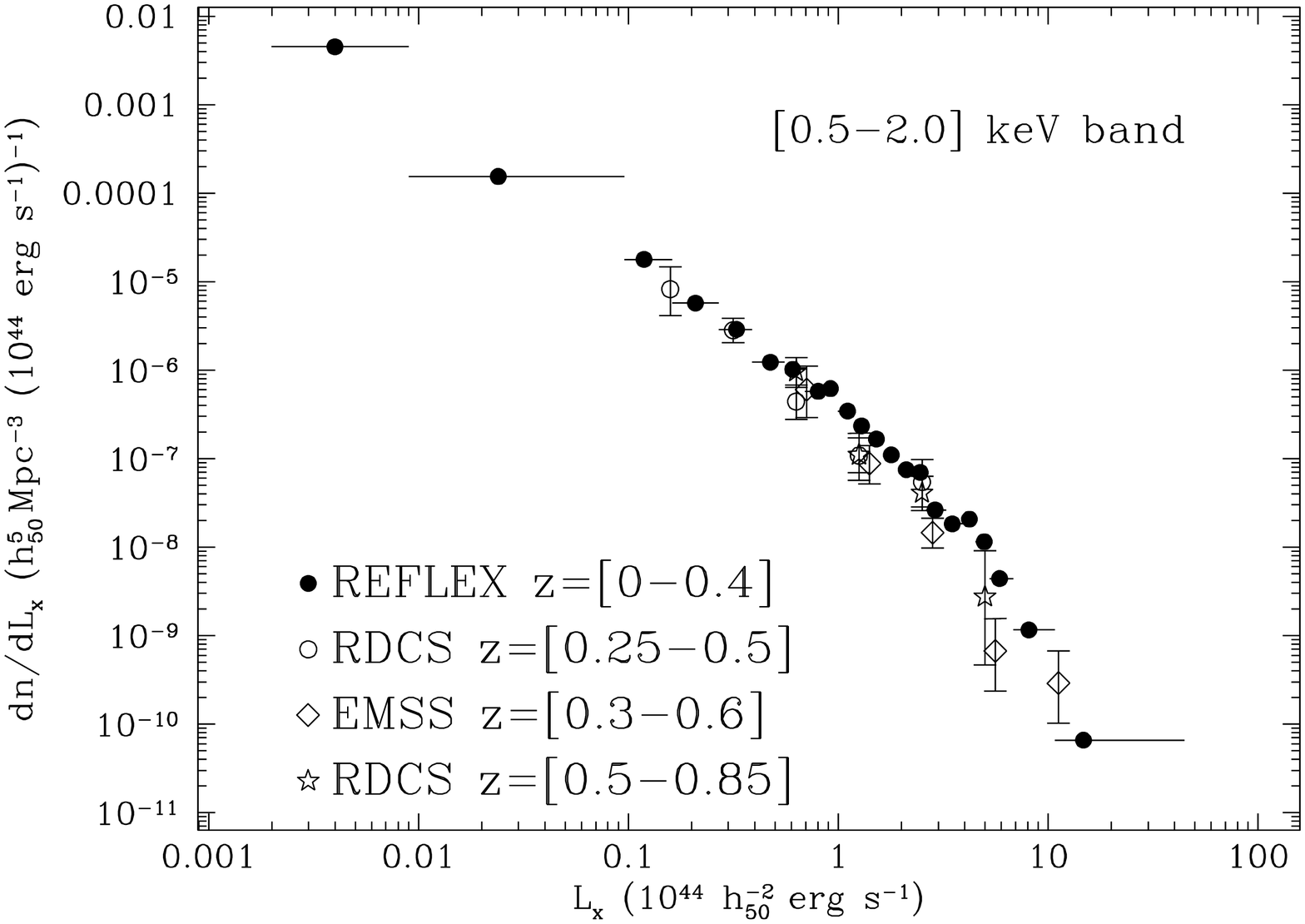}
\caption{Left: The REFLEX cluster X-ray luminosity function
(XLF, B\"ohringer et al. 2002), compared to the previous estimate obtained
from the RASS1 Bright Sample, the brighter forerunner of REFLEX (De
Grandi et al. 1999).  Right: Comparison of the REFLEX XLF, with that
of two representative distant  
cluster samples, the RDCS (Rosati et al. 1998) and EMSS (Henry et
al. 1992), showing the moderate reduction in the number density of
very luminous clusters at high redshifts.  The REFLEX XLF is here
computed in the [0.5--2.0] band, to allow for an
homogeneous comparison.}
\label{xlf}
\end{figure}

The X-ray luminosity function (XLF) of galaxy clusters is an excellent
observable 
incarnation of the cluster mass function, given that in clusters of
galaxies X-ray luminosity is well related to the cluster mass
(e.g. Reiprich \& B\"ohringer 2002).  For this reason, estimating the
XLF was one of the prime targets of the REFLEX survey.  It was also
one of the first results obtained by the project along the way, from
the higher--flux RASS1 Bright Sample (De Grandi et al. 1999).  The recent
XLF from the whole survey (B\"ohringer et al. 2002) is compared to
that early result in the left panel of Fig.~3.  The improvement in the
error bars of the 
latest REFLEX XLF (smaller than the size of the dots) gives an
immediate idea of its quality.  Such an 
accurate local measurement represents a standard 
reference of crucial importance for evolutionary studies, i.e. for
comparison with XLF measured at high redshift. One such example is
shown in the right panel of the same figure, where the 
REFLEX XLF is plotted together with those from the RDCS (Rosati et
al. 1998) and EMSS (Henry et al. 1992) deep surveys.  With the REFLEX
data pivoting the local abundance, the evidence for mild evolution of
the XLF bright end at high $z$ can be tested to a high accuracy
(e.g. Borgani et al. 2001; Henry 2001).
%
%

\section{The Clustering of REFLEX Clusters}

Clusters have a long history as tracers of large-scale structure (see
Nichol, this volume) and, in particular, surveys of X-ray clusters
have their own specific advantages (see Lahav et al. 1989 and Romer et
al. 1994 for early applications, and Borgani \& Guzzo 2001, Guzzo 2001
and Henry 2001 for recent reviews).  First of all, once the
characteristics of the X-ray instrument used are specified, a clean
selection function can be constructed.  This is by all means not an
obvious task for an optically-selected survey.  Second, fairly
precise predictions can be made from the models for both the mass
function and the clustering of objects above a given mass (see
e.g. Borgani, this volume and Moscardini et al. 2000).  These can be
reliably translated into ``observer space'' in terms of X-ray
luminosities and fluxes.  These advantages allow one to select samples
uniformly by mass over large areas (avoiding also other effects that
plague optically-selected samples as line-of-sight projections), and
provide the 
baseline for the REFLEX survey clustering results.

\begin{figure}
\plottwo{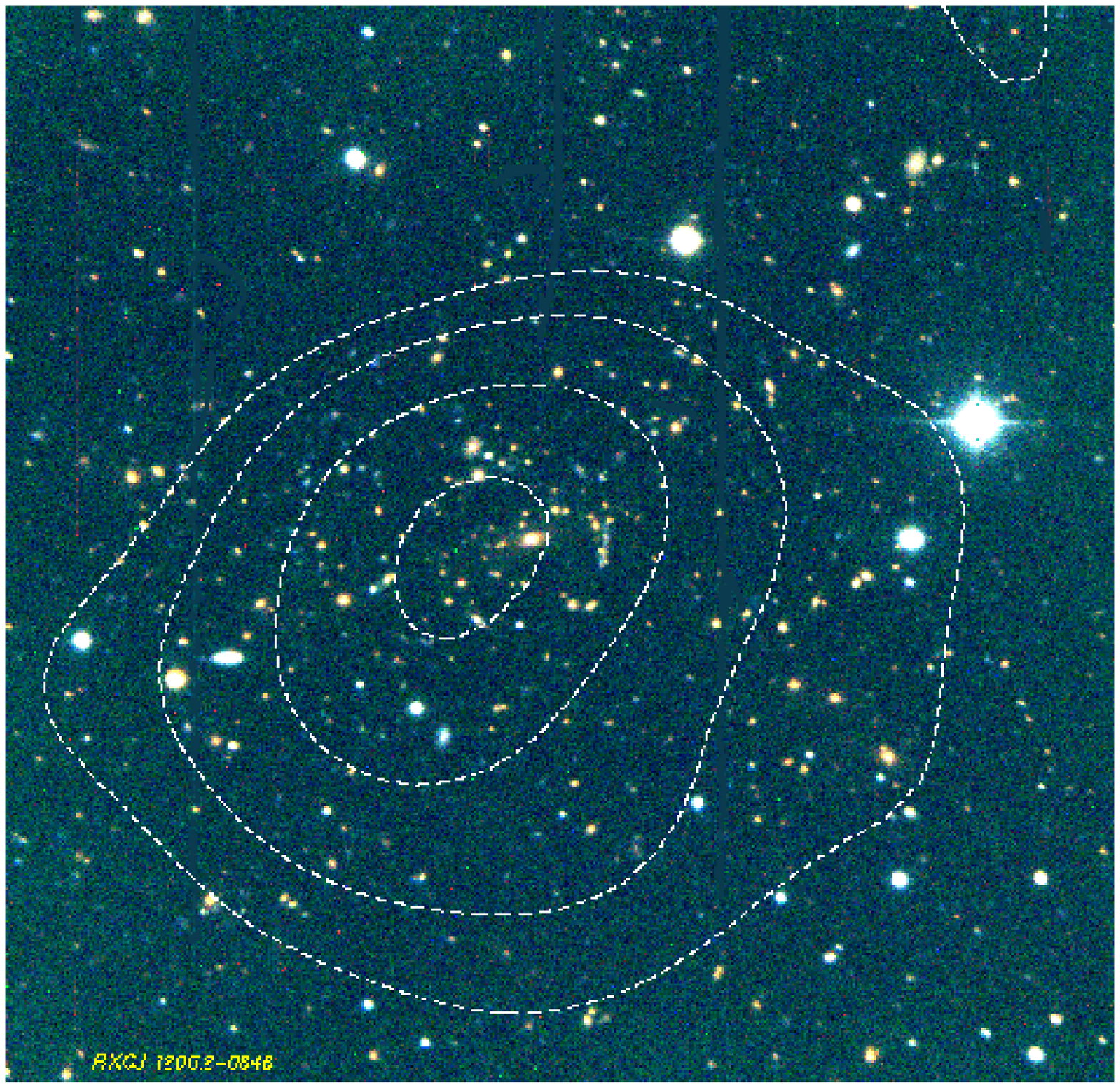}{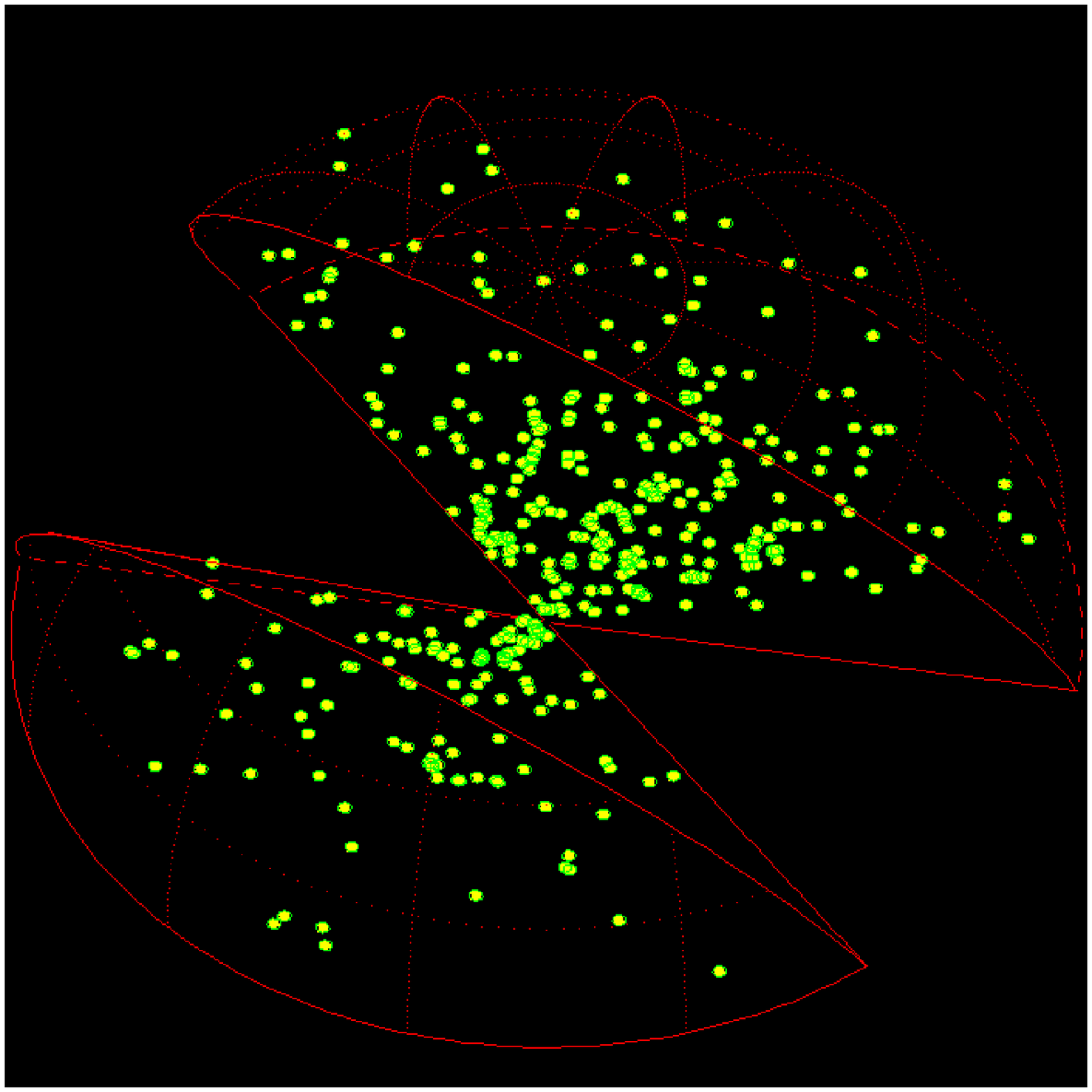}
\caption{Left: BVR combined mage of the spectacular cluster
RXCJ1206.2-0848, the second most distant object in the REFLEX survey,
at z=0.441. Right: The spatial distribution of REFLEX clusters, out to
600 h$^{-1}$ Mpc (plot from Borgani \& Guzzo 2001).  Note the presence
of agglomerates and ``chains'', demonstrating the strong
clustering of clusters among themselves.}
\label{fig:reflex_cone}
\end{figure}
The right panel of Fig.~4 plots the 3D distribution of REFLEX
clusters within a radius of 600 h$^{-1}$ Mpc.  Despite the fading with
distance due to the flux-limited selection function, a number of
superstructures with sizes 100 h$^{-1}$ Mpc are evident: clusters are
clearly still strongly clustered on such scales.  This can be
quantified by the two-point correlation function, that we plot for the
whole flux-limited survey in Fig.~5 (Collins et
al. 2000).  
\begin{figure}
\plotfiddle{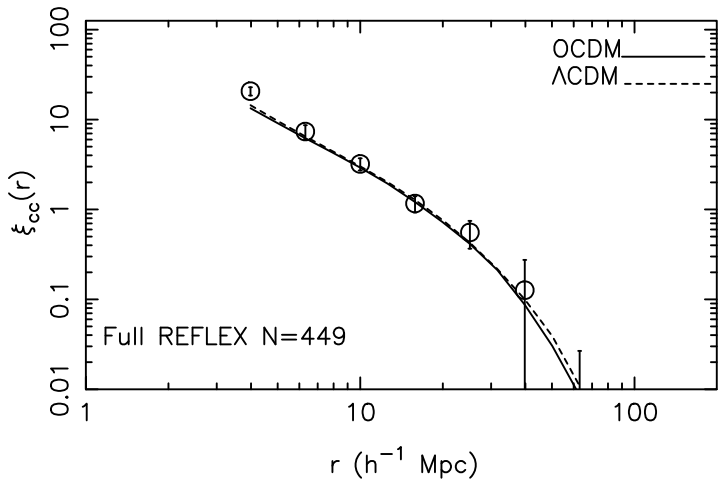}{6cm}{0}{40}{40}{-150}{210}
\caption{The REFLEX two-point correlation function compared with the
predictions for the same survey selection function of two
low-density CDM models, respectively open with $\Omega_m=0.3$,
$\Omega_{\Lambda}=0.0$ (OCDM) and flat with $\Omega_m=0.3$,
$\Omega_{\Lambda}=0.7$ ($\Lambda$CDM), as computed by
Moscardini et al. (2000).  (Here $r$ is the redshift-space separation,
usually called $s$ in galaxy surveys.)}  
\label{xi_reflex}
\end{figure}
\begin{figure}
\plotfiddle{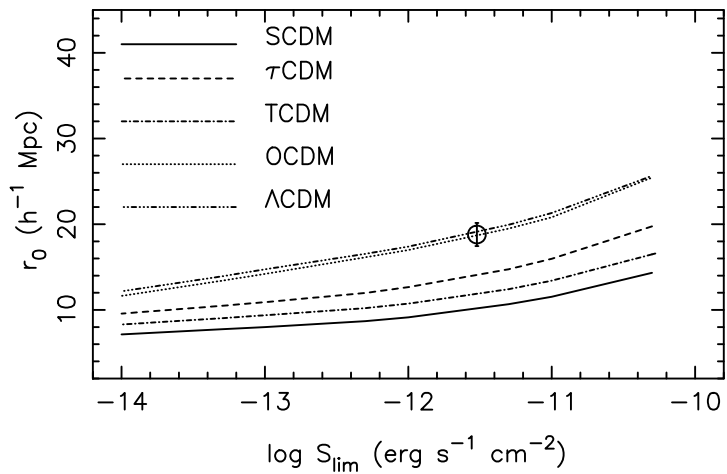}{6cm}{0}{40}{40}{-150}{210}
\caption{Comparison of the value of correlation length $r_0$ from the full
REFLEX flux-limited survey with the predictions as a function of
limiting X-ray flux for a range of CDM-type cosmological models
computed by Moscardini et al. (2000).}
\label{r0_fx}
\end{figure}
The shape of 
$\xi_{cc}$ is tightly constrained by the REFLEX data, and closely
reproduces an amplified version of the galaxy two-point correlation
function over almost two decades of scales, a classic prediction of
biasing theory 
(Kaiser 1984).  In the same figure, we also plot the predictions of
two low-density CDM models from Moscardini et al. (2000), which are
able to correctly reproduce both the shape and amplitude of the
observed REFLEX $\xi_{cc}$.  We remark that in this case the amplitude
is not a free parameter as for galaxy correlation functions, since the
bias value for the specific REFLEX selection function can be computed
using an appropriate theory for the clustering of massive haloes
(e.g. Mo \& White 1996 and subsequent refinements), given the fairly
straightforward relation between X-ray luminosity and mass (see Borgani \&
Guzzo 2001 for more details and references).  Analogous results have
been obtained from the power spectrum, as 
discussed in the contribution by P. Schuecker to this volume (and in
detail by Schuecker et al. 2001).   
\begin{figure}
\plotfiddle{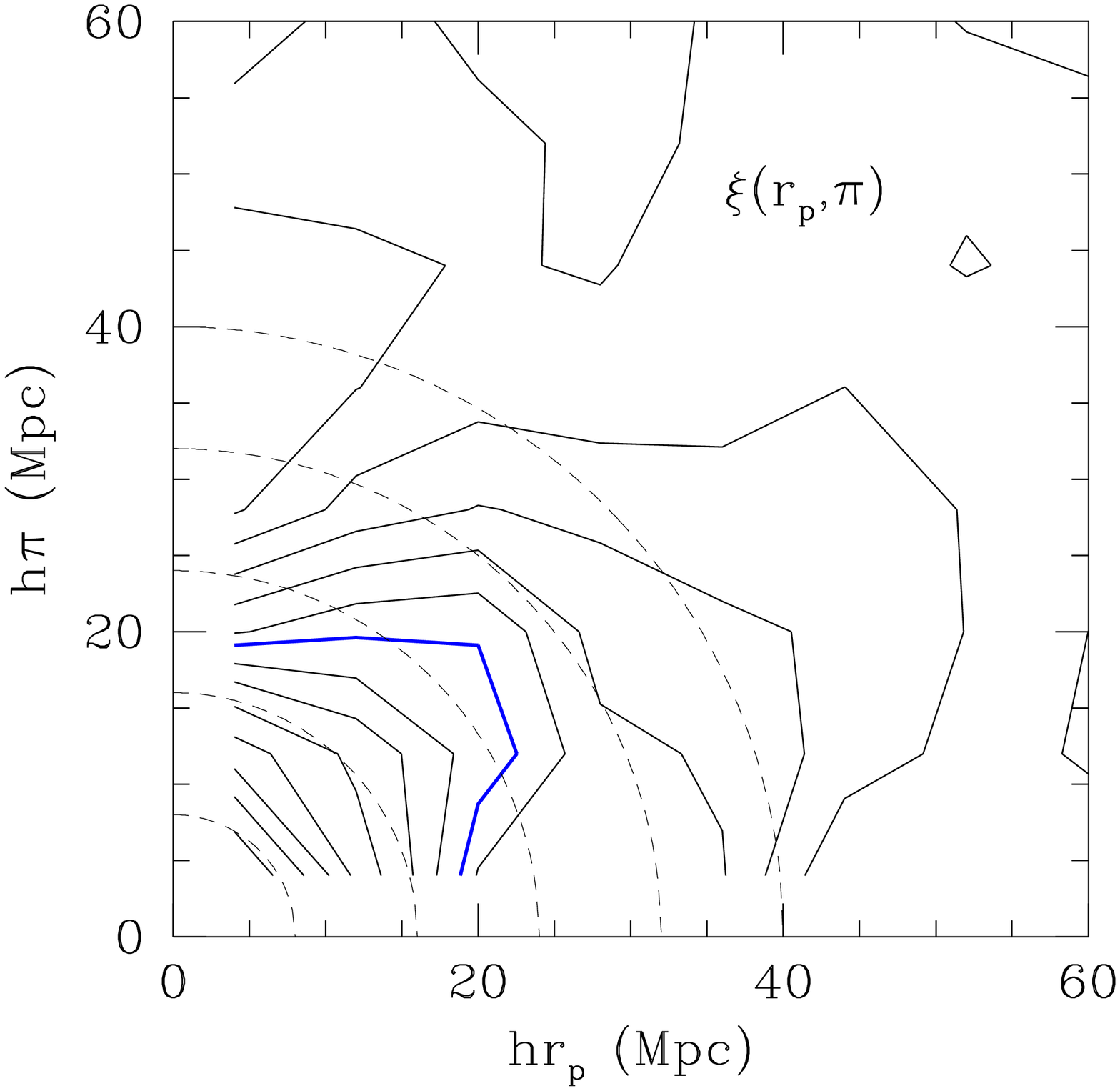}{6.5cm}{0}{35}{35}{-120}{-60}
\caption{The bi-dimensional correlation function $\xi(r_p,\pi)$, used 
to evidence redshift-space anisotropies in the clustering of REFLEX
clusters.  The diagram shows no stretching of the contours along the
line of sight, indicating that projection biases and redshift
errors are small in the REFLEX survey.  On the contrary, the
evident large-scale compression is a possible fingerprint of coherent
motions.  The dashed circles show how a perfectly isotropic
distribution would look like. } 
\label{csipz}
\end{figure}
Another prediction which can be precisely made for a survey like
REFLEX is the dependence of the correlation length on the survey flux
limit.  The expectation values for different CDM variants are
compared to the observed correlation length from the full REFLEX in
Fig.~6, again indicating that low-$\Omega_M$ models provide
the best consistency with observations (Collins et al. 2000).


Redshift-space anisotropies in the clustering pattern of clusters 
can be evidenced through the bidimensional
correlation function $\xi(r_p,\pi)$ [or analogously $\xi(\sigma,\pi)$]. 
Classical analyses of Abell samples show strong elongations along the 
line-of-sight $\pi$ direction (e.g. Miller et al. 2001), which are
typically not seen in automatic cluster catalogues (e.g. Nichol et
al. 1992).  These arise by a combination of projection effects, 
redshift errors and true spatial anisotropies in small--size
catalogues. The plot of $\xi(r_p,\pi)$ for REFLEX (Fig.~7),
in fact shows no evidence for stretching along the line 
of sight, indicating how these spurious effects are negligible
in this survey.  
On the other hand, the contours are significantly compressed at large
separations, a typical signature of streaming motions (proportional to
the quantity $\beta=\Omega_M^{0.6}/b$) which is for the
first time seen in a cluster sample (Guzzo et al. 2002; see also
Padilla \& Baugh 2001). 

We would like to conclude by mentioning that the statistical quality of
the results presented here will significantly improve in the near future,
thanks to the extension of REFLEX to fainter fluxes (by a factor $\sim
2$).  The REFLEX-2 sample is in fact being constructed and will bring
the total number of clusters over the same area to $\sim 800$ (see
e.g. B\"ohringer et al. 2001b).


\begin{references}

\reference Abell, G.O., 1958, \apjs, 3, 211
\reference Abell, G.O., Corwin, H.G. Jr., \& Olowin, R.P., 1989,
\apjs, 70, 1
\reference B\"ohringer, H., et al. (the REFLEX Team), 2001a, \aap, 369, 826
\reference B\"ohringer, H., et al. (the REFLEX Team), 2001b, {\it The
Messenger}, 106, in press
\reference B\"ohringer et al. (the REFLEX Team), 2002, \apj, in press
(astro-ph/0106243)
\reference Borgani S., \& Guzzo L., 2001, {\it Nature}, 409, 39
\reference Borgani, S., et al., 2001, \apj, in press (Nov 2001 issue,
also astro-ph/0106428)
\reference Collins, C.A., et al. (the REFLEX Team), 2000, \mnras, 319, 939
\reference De Grandi, S., et al. (the REFLEX Team), 1999, \apj, 513, L17 
\reference Guzzo, L. 2001, in {\it Where's the Matter?
Tracing Dark and Bright Matter with the New Generation of Large Scale
Surveys}, proc. of meeting held in Marseille (June 2001), M. Treyer \&
L. Tresse eds., Frontier Group (astro-ph/0111309)
\reference Guzzo, L., et al. (the REFLEX Team), 1999, {\it The
Messenger}, 95, 27
\reference Guzzo, L., et al. (the REFLEX Team), 2002, in preparation
\reference Henry, J.P., et al., 1992, \apj, 386, 408
\reference Henry, J.P., 2001, in ``AMiBA 2001: High-z
Clusters, Missing Baryons, and CMB Polarization'', in press
(astro-ph/0109498)
\reference Kaiser, N., 1984, \apj, 284, L9
\reference Lahav, O., et al., 1989, \mnras, 238, 881
\reference Miller, C.J., Nichol, R.C., \& Batuski, D.J., 2001, \apj,
555, 68
\reference Mo, H.J. \& White S.D.M., 1996, \mnras, 282, 347
\reference Moscardini, L., et al., 2000, \mnras, 314, 647
\reference Nichol, R.C., et al., 1992, \mnras, 255, 21p
\reference Padilla, N.D., Baugh, C.M., 2001, \mnras, submitted
(astro-ph/0104313) 
\reference Reiprich, T.H., \& B\"ohringer, H., 2002, \apj, in press
(astro-ph/0111285) 
\reference Romer, A.K., et al., 1994, {\it Nature}, 372, 75
\reference Rosati, P., et al., 1998, \apj, 492, L21
\reference Schindler, S.A., et al. (the REFLEX Team), 1995, \aap, 299, L9
\reference Schuecker et al. (the REFLEX Team), 2001, \aap, 368, 86
\end{references}
\end{document}